\documentclass[a4paper]{jpconf}
\usepackage{graphicx}
\usepackage{amsmath}
\usepackage{amssymb}
\begin{document}
\title{Chemical evolution of cold dark clouds in the vicinity of supernova remnants}

\author{A V Nesterenok}

\address{Ioffe Institute, 26 Polytechnicheskaya St., 194021, Saint Petersburg, Russia}

\ead{alex-n10@yandex.ru}

\begin{abstract}
The supernova explosion increases cosmic ray and x-ray fluxes in the surrounding interstellar medium. Cosmic ray particles and x-ray radiation penetrate nearby molecular clouds and affect the chemical and thermal evolution of the gas. Here we study chemical changes in the dense molecular gas influenced by a sudden increase of the ionization rate that may be caused by the supernova explosion. At the cloud core density $2 \times 10^4$~cm$^{-3}$, the H atom abundance reaches the equilibrium value at about $3 \times 10^4$~yr after the change in irradiation conditions. The response time of abundances of icy mantle species is $10^4-10^5$~yr. The abundances of neutral and grain mantle species may not reach their equilibrium values in molecular clouds in the vicinity of 'middle-aged' supernova remnants.
\end{abstract}

\section{Introduction}
In cold dark clouds, cosmic rays are the primary ionization agents, since interstellar ultraviolet radiation (UV) is almost absorbed in the surface layer of the cloud of 3--4 mag visual extinction \cite{Hollenbach2009}. The ion--molecule chemistry begins with the ionization of H$_2$ by cosmic rays and fast ion--molecule reaction between H$^+_2$ and H$_2$. As a result, H$_3^+$ ions are formed that subsequently react with the abundant neutral species (CO, N$_2$, H$_2$O, HD and etc.) \cite{Larsson2012}. Ion--molecule reactions play a central role in the production of polyatomic species in the cold interstellar gas and are important heating mechanism at high visual extinctions \cite{Glassgold2012}.

Cosmic ray interactions with the interstellar gas produce energetic electrons that collisionally excite electronic states of H$_2$ molecules. The subsequent radiative de-excitations of H$_2$ generate a flux of far-ultraviolet photons that dissociate and ionize chemical species \cite{Heays2017}. Other effect of non-thermal electrons is the gas heating through the rotational excitation of H$_2$ molecules (and subsequent collisional de-excitation), Coulomb losses and momentum transfer with neutral particles \cite{Dalgarno1999}.

It is possible to estimate the cosmic ray ionization rate based on the relative abundances of different molecular tracers \cite{Dalgarno2006}. The ionization rate in the dense molecular gas shielded from the sources of ionizing radiation (such as newly-formed stars, supernova remnants) is found to lie in the range $10^{-17}-10^{-16}$~s$^{-1}$ \cite{Caselli1998,Williams1998}. However, the proximity of sources of the ionizing radiation such as supernovae may significantly enhance fluxes of cosmic rays and x-ray radiation. For example, the cosmic ray ionisation rate was found to be about 10$^{-15}$~s$^{-1}$ towards positions located close to the supernova remnant W28 \cite{Vaupre2014} and near the W51C \cite{Ceccarelli2011,Shingledecker2016}. Here we study the response of the chemical composition of the dense interstellar gas to a sudden change in irradiation conditions that may be caused by the supernova explosion.

\section{Description of the model}
A simple zero-dimensional model is considered in which the gas density remains fixed as the chemistry develops out some initial state. We take into account the gas-phase chemistry, the adsorption of gas species on dust grains, various desorption mechanisms, the grain surface chemistry, the ion neutralization on dust grains -- the details of the calculations can be found in \cite{Nesterenok2018}. The gas-phase chemical network is based on the UMIST Database for Astrochemistry (UDfA), 2012 edition \cite{McElroy2013}. The grain surface network is based on the NAUTILUS code network \cite{Ruaud2016}. Branching ratios for the reactions involving carbon-chain species given by \cite{Chabot2013} are used. Photodissociation and photoionisation rates of chemical species by cosmic-ray induced UV flux are updated according to \cite{Heays2017}. The dissociation rate of molecular hydrogen by cosmic ray particles was taken according to the recent study by \cite{Padovani2018}. We use solid/gas photodissociation coefficient ratio equal to 0.1 \cite{Kalvans2018}. Parameters of the grain surface chemistry adopted in simulations are listed in the table~3 in \cite{Nesterenok2018}.

A single-size grain model is considered, the grain radius is 0.05~$\mu$m and the dust--gas mass ratio is equal to 0.01. Corresponding grain surface area is about $10^{-21}$~cm$^{2}$~per~H. The dust temperature is taken equal to 10~K. The gas temperature is obtained by solving equation of thermal balance where main processes of the gas heating and cooling are taken into account \cite{Nesterenok2018}. 

The population densities of energy levels of ions CI, CII and OI and molecules H$_2$, CO, H$_2$O are computed. The details on spectroscopic data and data on collisional rate coefficients used in simulations are given by \cite{Nesterenok2018}. We include the collisional excitation of molecular hydrogen by non-thermal electrons produced by the cosmic ray ionization \cite{Tine1997}.

The parameters of the dark cloud core adopted in simulations are given in the table~\ref{table}. Here we assume that the ionization of the gas is governed by one parameter -- ionization rate $\zeta_{\rm{H_2}}$ -- that is attributed to the interaction of cosmic rays with the medium. We note that cosmic rays and x-rays affect the molecular gas similarly \cite{Mackey2019}. Our definition of the cosmic ray ionization rate is related to the definition of the parameter used in the UDfA \cite{McElroy2013}. The rate of the electron production in the unit volume by the ionization of molecular gas is given by $\zeta_{\rm{H_2}} n_{\rm{H_2}}$, where He ionization is taken into account. The initial value of the cosmic ray ionization rate is set equal to $\zeta_{\rm{H_2,0}} = 10^{-17}$~s$^{-1}$. The elemental abundances used in our calculations are low-metal abundances, the C/O ratio is equal to 0.5, see table~1 in \cite{Nesterenok2018}. The hydrogen is set to be molecular and all the other elements are set to be in atomic form initially. The chemical composition of the gas and icy mantles of dust grains is calculated until the evolution time 0.5~Myr (the typical lifetime of dense cloud cores is of the order of 1~Myr \cite{Andre2014}). At this time, three scenarios are considered: the cosmic ray ionization rate is remained unchanged, and the cosmic ray ionization rate is abruptly increased up to a new value $\zeta_{\rm{H_2}}$, two values are considered -- $10^{-16}$ and $10^{-15}$~s$^{-1}$. 

\begin{table}
\caption{\label{table}Parameters of the dark cloud core.}
\begin{center}
\begin{tabular}{ll}
\br
Gas density, $n_{\rm{H, tot}}$ & $2 \times 10^4$~cm$^{-3}$\\
Visual extinction, $A_{\rm{V}}$ & 10 \\
Micro-turbulence speed, $v_{\rm{turb}}$ & 0.3~km~s$^{-1}$ \\
Initial cosmic ray ionization rate, $\zeta_{\rm{H_2,0}}$ & $10^{-17}$~s$^{-1}$ \\
Dust temperature, $T_{\rm{d}}$ & 10~K \\
Grain surface area density & 9.7$\times 10^{-22}$~cm$^{2}$~per~H \\
\br
\end{tabular}
\end{center}
\end{table}

\section{Results}

\begin{figure}[h]
\includegraphics[width=9cm]{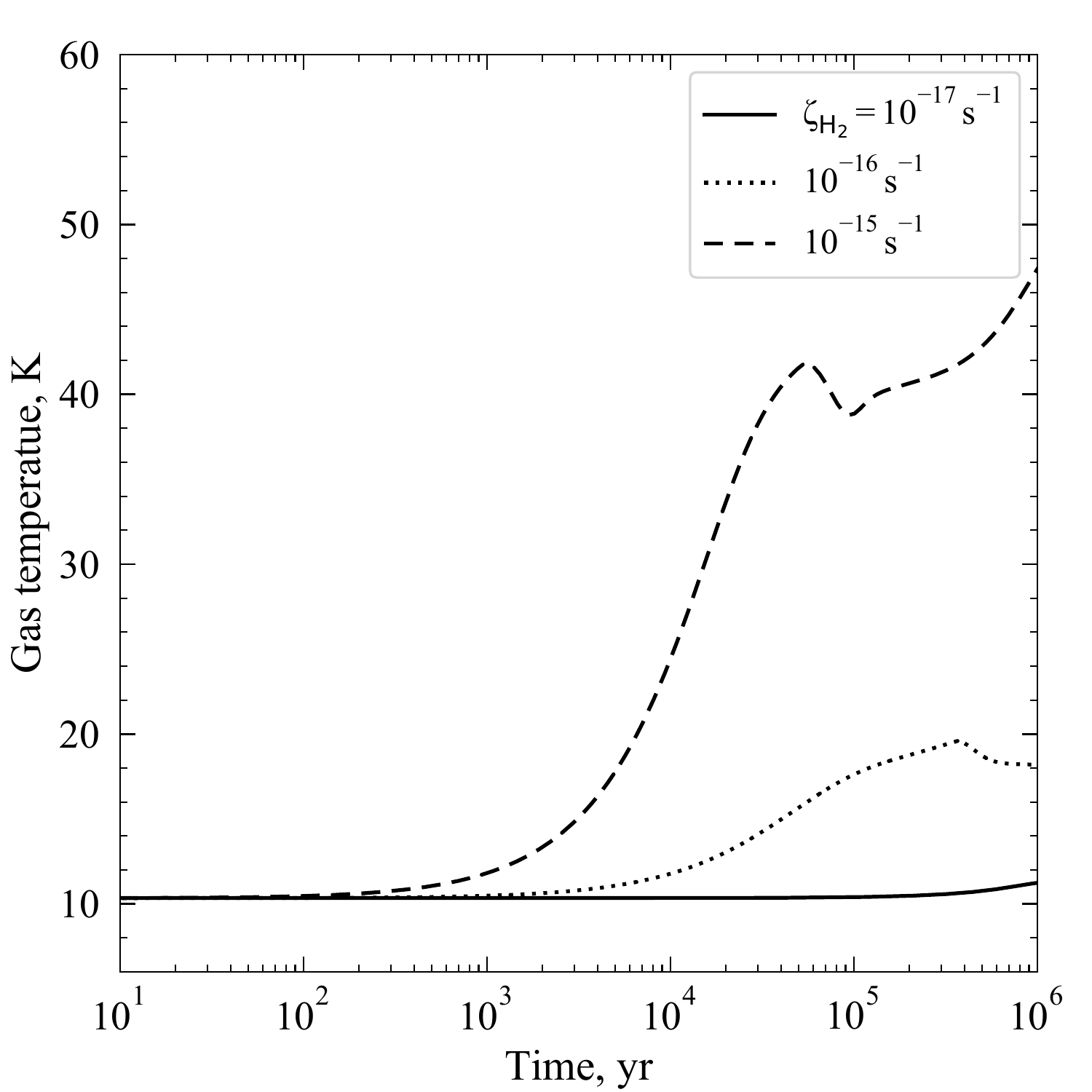}\hspace{2pc}
\begin{minipage}[b]{6cm}\caption{\label{fig_gas_temp}The gas temperature as a function of time passed after the increase of the ionization rate. Three scenarios are considered -- the cosmic ray ionization rate is remained unchanged (solid line), the cosmic ray ionization rate is increased up to a new value, $10^{-16}$~s$^{-1}$ (dotted line) and 10$^{-15}$~s$^{-1}$ (dashed line).}
\end{minipage}
\end{figure}

The temperature of the neutral gas component is shown on the figure~\ref{fig_gas_temp}. At $\zeta_{\rm{H_2}} = 10^{-15}$~s$^{-1}$, the gas reaches a thermal equilibrium at about few $10^4$~yr after the increase of the cosmic ray ionization rate. The cosmic ray driven chemistry and the thermalization of secondary energetic electrons are both important in heating the gas. The H$_2$ molecules are ro-vibrationally excited by non-thermal electrons, and the H$_2$ de-excitation in collisions with neutral species heats the gas. The main cooling mechanism is the cooling via line emission by atoms and molecules at 10~K. At higher gas temperature, the gas cooling via gas--dust collisions becomes important. Dissociative recombination reactions and many ion--neutral reactions are faster at low temperature \cite{McElroy2013}. It may have some effect on the abundance of neutral species that are destroyed in the gas phase by ion--neutral reactions. However, the main changes in the chemical evolution are induced by the increase of the ionization fraction of the gas. At $\zeta_{\rm{H_2}} = 10^{-16}$~s$^{-1}$, the response of the chemistry and the gas temperature to the change of ionization rate is slower than at $10^{-15}$~s$^{-1}$.

\begin{figure}[h]
\begin{center}
\includegraphics[width=16.5cm]{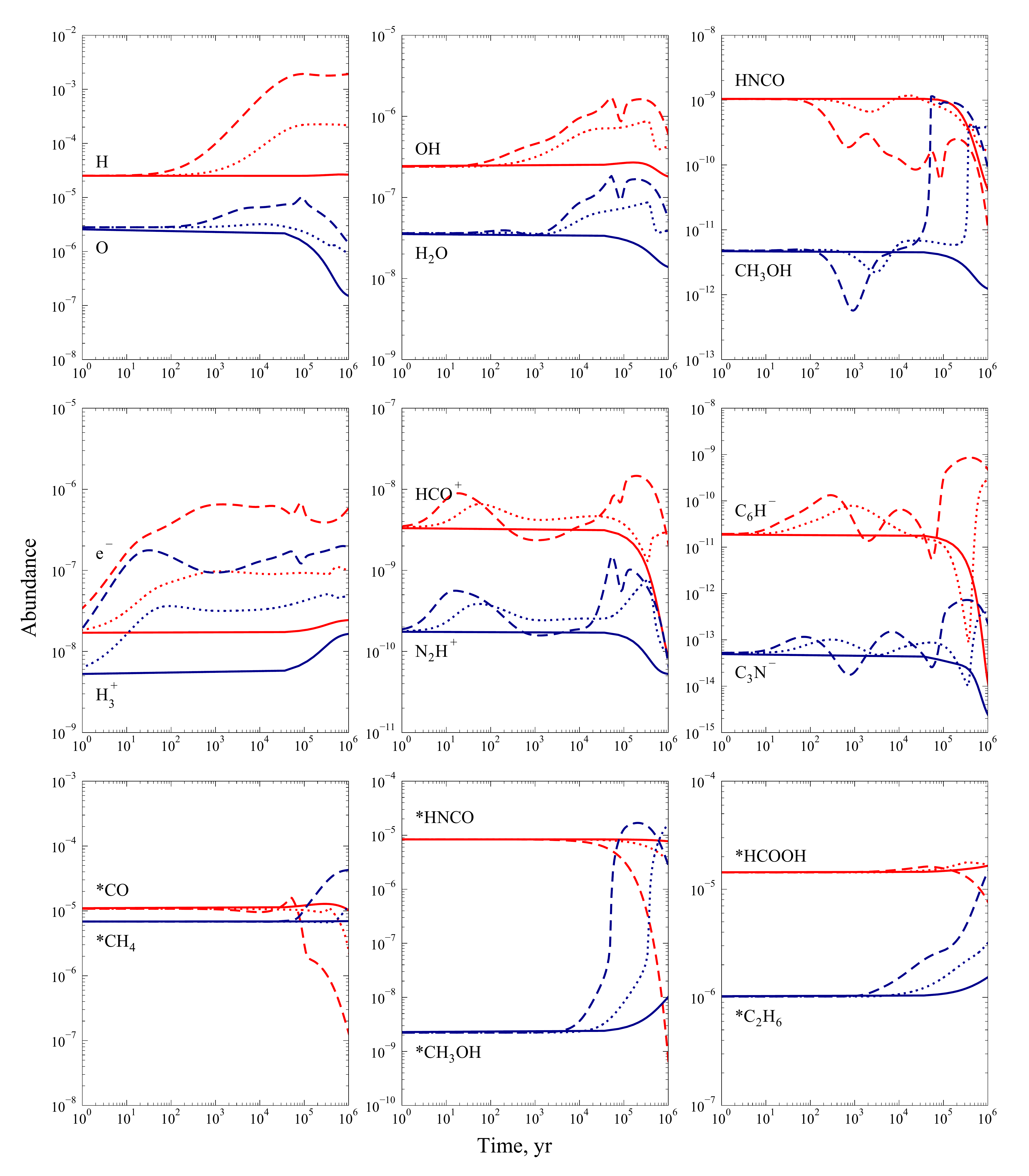}
\caption{Abundances of chemical species in the dark cloud core as a function of time passed after a sudden increase of the ionization rate. The upper row of plots -- gas-phase species, the middle -- ions, lower -- species adsorbed on dust grains. Solid line corresponds to the constant ionization rate of 10$^{-17}$~s$^{-1}$ during entire evolution time, dotted line corresponds to the final cosmic ray ionization rate of 10$^{-16}$~s$^{-1}$, dashed line -- 10$^{-15}$~s$^{-1}$.}
\label{fig_chem_ev}
\end{center}
\end{figure}

Figure~\ref{fig_chem_ev} shows fractional abundances of selected species relative to the total hydrogen nuclei number density as a function of time elapsed after the change of irradiation conditions. Ions that are produced directly by the cosmic ray ionization have the shortest time of reaching steady state. The higher the ionization rate, the higher the abundance of H$^+$, H$_2^+$, H$_3^+$, He$^+$. The H$_3^+$ abundance increases up to about few 10~yr and after remains approximately constant -- the ion production rate depends on the cosmic ray ionization rate and is constant, but the destruction rate increases with increasing electron abundance. The abundances of secondary ions such as HCO$^+$ and N$_2$H$^+$ are determined by the rates of formation reactions involving H$_3^+$ and reactions of dissociative recombination. Anions C$_6$H$^-$ and C$_3$N$^-$ are mostly produced by electron attachment reactions, and are destroyed in reactions with positive ions and H atoms. The evolution of abundance of anions reflects the abundance of parent species -- C$_6$H, C$_3$N, HNC$_3$.

Neutral species -- H, OH, H$_2$O --  reach their new abundances at evolutionary times $t \sim 10^3-10^4$~yr after the cosmic ray ionization increase. Both OH, H$_2$O are produced in the gas phase in the chain of ion--neutral reactions starting with atomic oxygen. H$_2$O is also produced on the surface of dust grains with subsequent ejection to the gas phase through chemical desorption. OH production by photo-dissociation of water is found to be not important in agreement with \cite{Farquhar1994}. The main destruction channel of HNCO and CH$_3$OH in the gas phase are reactions with ions. The abundances of these species drop as the ionization fraction increases. At later times, the gas-phase abundance of methanol increases by two orders of magnitude due to the chemical desorption from the surface of dust grains. However, the abundance of methanol in the gas phase remains relatively low -- about 10$^{-9}$.

The competition between the destruction of H$_2$ by cosmic rays and the H$_2$ formation on dust grains determines the abundance of H atoms in the gas phase. The time needed to achieve the equilibrium abundance for H atoms does not depend on the cosmic ray ionization rate but is inversely proportional to the rate of H$_2$ formation on dust grains. Once the H atom adsorbs on the grain surface, it quickly reacts with one of icy species (CO, HCO, NO, HNO and others), and, as a result, H$_2$ molecules form through hydrogen abstraction reactions. The formation rate of H$_2$ can be approximately expressed as $k = R_{\rm{H_2}} \, n_{\rm{H, tot}} \, n_{\rm{H}}$, where effective rate coefficient $R_{\rm{H_2}} \simeq (2-3) \times 10^{-17}$~cm$^{3}$~s$^{-1}$ in our model. The coefficient $R_{\rm{H_2}}$ depends on the adsorption rate of H atoms, which in turn depends on the gas temperature.

The increase of the cosmic ray ionization rate leads to the increase of the H atom abundance in the gas phase and on the grain surface. The elevated abundance of H atoms on the surface of dust grains initiates the chain of reactions:

\begin{equation}
\rm{CO \to HCO \to H_2CO \to CH_2OH \,\, or \,\, CH_3O \to CH_3OH,} 
\end{equation}

\noindent
that eventually convert adsorbed CO into methanol. This chain of reactions is accompanied by desorption of species due to the exothermicity of the reactions. The reactive desorption fraction -- the efficiency of product desorption from the surface after reaction -- is taken equal 0.01 in our model \cite{Chuang2018}. Abundances of other oxygen bearing species (HNCO, HCOOH) on the surface of dust grains diminish as they have CO in their production chains. The high abundance of H atoms shifts the chemistry to the formation of hydrocarbons such as methane CH$_4$, ethane C$_2$H$_6$, see figure~\ref{fig_chem_ev}. The effect of H atom abundance on the grain surface chemistry is also discussed by \cite{Cuppen2009}. The response time of the grain surface chemistry to the changes in irradiation conditions is $t > 10^4$~yr. 

Other chemical effect of high H atom abundance is the temporary twofold increase of CO abundance in the gas phase -- via reaction H + HCO $\to$ H$_2$ + CO, where HCO is ejected from icy mantles of dust grains through the chemical desorption. As a result, the gas cooling becomes more efficient and the gas temperature decreases by a few K at about $(0.5-1) \times 10^5$~yr, see figure~\ref{fig_gas_temp}. At later evolutionary times, the gas phase abundance of CO diminishes due to freeze-out onto dust grains, the gas temperature increases.

\section{Discussions and conclusions}

The effect of elevated cosmic ray ionization rate on the interstellar gas chemistry was considered by a number of authors, e.g. \cite{Shingledecker2016,Farquhar1994,Albertsson2018,Ceccarelli2018}. The important task is to find molecular tracers that can be used to determine the cosmic ray ionization rate based on observational data. Usually, a fixed value of the ionization rate that does not change during the chemical evolution is considered in simulations. Chemical changes caused by a sudden cosmic ray irradiation are dominated by ion--neutral reactions that are generally free of activation energy barriers and have rates comparable with collisional rates \cite{Black1998}. In this case, it is usually assumed that time-scale to achieve steady state is short, e.g. \cite{Becker2011}. Indeed, the response time of ionization fraction is $t \sim 10-10^2$~yr, see figure~\ref{fig_chem_ev}. However, neutral species reach their equilibrium abundances at larger times, $t \gtrsim 10^3$~yr. The H/H$_2$ ratio reaches the equilibrium value at about $3 \times 10^4$~yr after the change of irradiation conditions. Main changes in the grain surface chemistry are determined by the H atom abundance rise, and the response time is $10^4-10^5$~yr. Thus, the abundances of some neutral and grain mantle species may not reach their equilibrium values in molecular clouds in the vicinity of 'middle-aged' supernova remnants (with ages of the order of 10$^4$~yr), e.g. W28, W51C.

If a supernova remnant encounters molecular clumps of dense gas, the shock wave slows down. The non-dissociative (C-type) shock wave may develop where the gas is heated but nevertheless molecules survive the passage of the shock front \cite{Nesterenok2018}. The well-known examples of such interactions are W44, W28, IC 443, W51C \cite{Ceccarelli2011,Neufeld2007}. In the shock wave, ice mantles of dust grains are liberated into the gas phase, and molecules are chemically processed in the hot shocked gas. As a result, the composition of icy mantles of dust grains in the preshock cloud affects the gas phase molecular abundances in the postshock region. On the other hand, the H/H$_2$ ratio in the preshock gas determines the rate of para-to-ortho-H$_2$ conversion and the collisional excitation of vibrational states of H$_2$ in the shock wave \cite{Neufeld2007}. The results of this study are important in interpretation of observations of molecular emission from shocks in supernova remnants. 

\section*{References}
\bibliography{references} 

\end{document}